# Astro2020 Science White Paper

## *Dark Cosmology*: Investigating Dark Matter & Exotic Physics in the Dark Ages using the Redshifted 21-cm Global Spectrum

**Thematic Areas:** ☐ Planetary Systems  ☐ Star and Planet Formation
☐ Formation and Evolution of Compact Objects  ☒ Cosmology and Fundamental Physics
☐ Stars and Stellar Evolution  ☐ Resolved Stellar Populations and their Environments
☐ Galaxy Evolution  ☐ Multi-Messenger Astronomy and Astrophysics

**Principal Author:**
Name: Jack O. Burns
Institution: University of Colorado Boulder
Email: jack.burns@colorado.edu
Phone: 303-735-0963

**Co-authors:** S. Bale (UC-Berkeley), N. Bassett (U. Colorado), J. Bowman (ASU), R. Bradley (NRAO), A. Fialkov (U. Sussex), S. Furlanetto (UCLA), M. Hecht (Haystack Obs.), M. Klein-Wolt (Radbound U.), C. Lonsdale (Haystack Obs.), R. MacDowall (GSFC), J. Mirocha (McGill), J. Muñoz (Harvard), B. Nhan (U. Virginia), J. Pober (Brown), D. Rapetti (U. Colorado), A. Rogers (Haystack Obs.), K. Tauscher (U. Colorado)

**Abstract**: The Dark Ages, probed by the redshifted 21-cm signal, is the ideal epoch for a new rigorous test of the standard ΛCDM cosmological model. Divergences from that model would indicate new physics, such as dark matter decay (heating) or baryonic cooling beyond that expected from adiabatic expansion of the Universe.

After the Cosmic Microwave Background photons decoupled from baryons, the Dark Ages epoch began: density fluctuations grew under the influence of gravity, eventually collapsing into the first stars and galaxies during the subsequent Cosmic Dawn. In the early Universe, most of the baryonic matter was in the form of neutral hydrogen (HI), detectable via its ground state's "spin-flip" transition. A measurement of the redshifted 21-cm spectrum maps the history of the HI gas through the Dark Ages and Cosmic Dawn and up to the Epoch of Reionization (EoR), when ionization of HI extinguished the signal. The *Experiment to Detect the Global EoR Signature* (EDGES) recently reported an absorption trough at 78 MHz (redshift z~17), similar in frequency to expectations for Cosmic Dawn, but ~3 times deeper than was thought possible from standard cosmology and adiabatic cooling of HI. Interactions between baryons and slightly-charged dark matter particles with electron-like mass provide a potential explanation of this difference but other cooling mechanisms are also being investigated to explain these results.

The Cosmic Dawn trough is affected by cosmology and the complex astrophysical history of the first luminous objects. Another trough is expected during the Dark Ages, prior to the formation of the first stars and thus determined entirely by cosmological phenomena (including dark matter). Observations on or in orbit above the Moon's farside can investigate this pristine epoch (~15-40 MHz; z~100-35), which is inaccessible from Earth. A single cross-dipole antenna or compact array can measure the amplitude of the 21-cm spectrum to the level required to distinguish (at >5$\sigma$) the standard cosmological model from that of additional cooling derived from current EDGES results. In addition to dark matter properties such as annihilation, decay, temperature, and interactions, the low-frequency background radiation level can significantly modify this trough. Hence, this observation constitutes a powerful, clean probe of exotic physics in the Dark Ages.

# Introduction

After the Big Bang, the Universe was hot, dense, and nearly homogeneous. As the Universe expanded, the material cooled, condensing after ~400,000 years (z~1100) into neutral atoms, freeing the cosmic microwave background (CMB). The baryonic content of the Universe consisted primarily of neutral hydrogen. Fifty million years or so later, gravity drove the formation of the first luminous objects – stars, black holes, and galaxies – which ended the Dark Ages and initiated the Cosmic Dawn (z∼20-30; see e.g., Loeb & Furlanetto 2013). These first stars likely differed dramatically from nearby stars, as they formed in vastly different environments (e.g., Abel et al. 2002). Figure 1 places this epoch into perspective.

This transformative time period marked the first emergence of complexity in our Universe, **but no currently-planned telescope can explore the highest redshifts (z≳30) of this epoch.** While JWST, WFIRST, and a suite of ground-based facilities will observe the Universe as it was ~300 million years after the Big Bang (and especially focus on the Reionization era, when distant galaxies ionized the gas between them about a billion years after the Big Bang), none now contemplate observing the true first stars and black holes (e.g., Behroozi & Silk 2015) much less the Dark Ages that preceded them. For example, CMB observations of Thomson scattering measure the integrated column density of ionized hydrogen, but only roughly constrain the evolution of the intergalactic medium (e.g., Mesinger et al. 2012). Ly-$\alpha$ absorption from QSOs only constrains the end of reionization at relatively late times, z∼7 (e.g., McGreer et al. 2015). Observations with HST and JWST will only find the brightest galaxies at high redshifts (z≲15), and thus any inferences drawn about the high-z galaxy population as a whole depend upon highly-uncertain assumptions (e.g., Bouwens et al. 2015). The Hydrogen Epoch of Reionization Array (HERA; DeBoer et al. 2017) will observe the neutral hydrogen (HI) power spectrum from ∼50-200 MHz (z=27-6) probing the Cosmic Dawn and EoR epochs but not the Dark Ages (e.g., Ewall-Wice et al. 2016). On the other hand, space-based observations at frequencies ≲40 MHz using the 21-cm spin-flip transition of HI can investigate the Global Signal from the pre-stellar epoch at z∼50-100 which is not accessible from the ground.

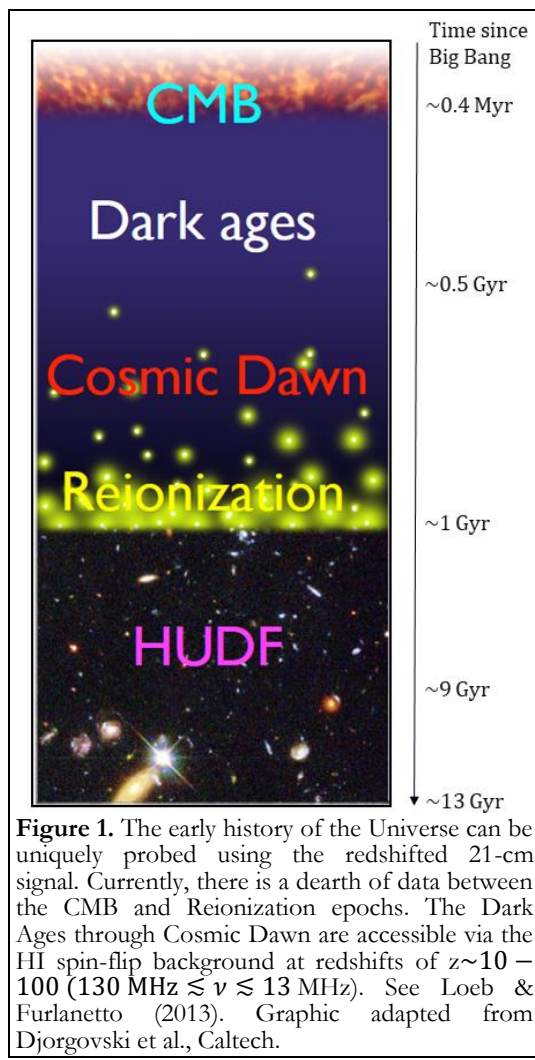

**Figure 1.** The early history of the Universe can be uniquely probed using the redshifted 21-cm signal. Currently, there is a dearth of data between the CMB and Reionization epochs. The Dark Ages through Cosmic Dawn are accessible via the HI spin-flip background at redshifts of z∼10 − 100 (130 MHz ≲ $\nu$ ≲ 13 MHz). See Loeb & Furlanetto (2013). Graphic adapted from Djorgovski et al., Caltech.

In this white paper, we describe the breakthrough science that can be performed by probing the Dark Ages for the first time using the highly redshifted 21-cm monopole. An absorption trough expected below 40 MHz is sensitive to deviations from the standard ΛCDM cosmology model generated, for example, by added heating or cooling possibly produced by dark matter. Such very low frequency radio astronomy observations must be performed from the lunar farside to escape Earth-based ionospheric, radio-frequency interference (RFI), and ground effects. We show that a relatively simple orthogonal, dual-dipole antenna with a radio spectrometer/polarimeter, each with significant flight heritage, can probe for new physics beyond the standard model using the 21-cm global spectrum.

## The Redshifted 21-cm Global Spectrum

The most promising method to measure the properties of the Dark Ages is the highly-redshifted Global or All-sky 21-cm Spectrum illustrated in Figure 2 (see e.g., Pritchard & Loeb 2012). The "spin-flip" transition of neutral hydrogen emits a photon with a rest wavelength of 21-cm ($\nu$=1420 MHz). This signal, while generated by a weak hyperfine transition, is nevertheless observable because neutral hydrogen pervades the Universe during the Dark Ages and the onset of Cosmic Dawn.

The curves in Figure 2 illustrate broad spectral features that are common to virtually all 21-cm models of the early Universe. The evolution of the brightness temperature is driven by the evolution of the ionization fraction ($x_{HI}$) and spin temperature $T_S$ (measure of the fraction of atoms in the two spin states) of HI relative to the radio background temperature $T_R$ (usually assumed to be the CMB) (e.g., Furlanetto et al. 2006; Shaver et al. 1999) as given by

$$\delta T_b \simeq 27\, \bar{x}_{H\,I}(1+\delta) \left(\frac{\Omega_{b,0} h^2}{0.023}\right) \left(\frac{0.15}{\Omega_{m,0} h^2} \frac{1+z}{10}\right)^{1/2} \left(1 - \frac{T_R}{T_S}\right) \text{ mK} \qquad \text{Eq. (1)}$$

where $\delta$ is the baryon overdensity (∼0 for the all-sky signal), $h$ is the normalized Hubble constant, and $\Omega_b$ ($\Omega_m$) is the baryon (total mass) abundance parameter. Because of cosmic expansion, the observed frequency $\nu_0$ corresponds to a redshift z through the relation $\nu/\nu_0$=1+z, where $\nu$=1420 MHz is the rest frequency; so, the 21-cm spectrum is a powerful measure of the time evolution of structure growth in the early Universe (see Figure 2).

In the standard ΛCDM cosmological model, the lowest frequency (corresponding to the highest redshift and earliest time) spectral absorption feature in each curve of Figure 2, hereafter called the "Dark Ages trough" ($\nu$<40 MHz), is purely cosmological, containing no information about the history or astrophysics of stars and galaxies. At z≳30, cosmic expansion drives a decoupling between the spin temperature and the radio background radiation temperature ($T_R$ > $T_S$) producing a broad absorption feature in the 21-cm spectrum. The standard cosmological model makes a precise prediction of the central frequency (≈18 MHz) and brightness temperature (≈40 mK) for this feature. Any departure from these values (e.g., the red curve in Figure 2) would indicate a deviation from the standard model, with additional exotic physics such as non-gravitational interactions between baryons and dark matter being required as recently suggested (e.g., Barkana 2018; Muñoz & Loeb 2018; Berlin et al. 2018; Muñoz et al. 2018). Thus, **the low frequency 21-cm spectrum offers a novel and powerful probe of the standard cosmological model.**

Recent results (Bowman et al. 2018a,b) reported by the *Experiment to Detect the Global Epoch of Reionization (EoR) Signature* (EDGES), suggest the presence of a strong absorption feature in the 21-cm spectrum at ≈78 MHz (z≈17), within the range expected for the "Cosmic Dawn trough" caused by the onset of the first stars/galaxies/black holes. If verified by other experiments, these results offer the exciting prospect of investigating physics outside the standard cosmological model in the early Universe.

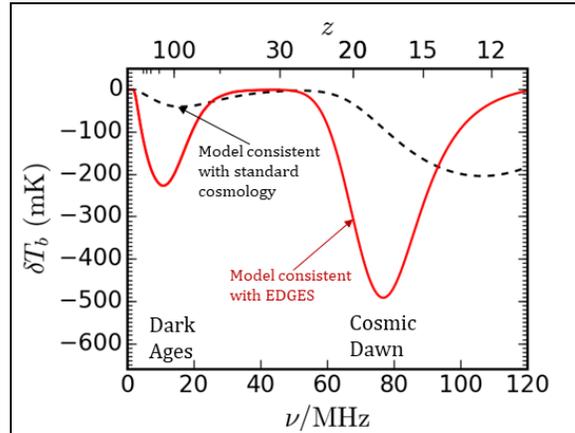

**Figure 2**. The redshifted Global 21-cm spectrum provides crucial cosmological information from the Dark Ages through Cosmic Dawn. The curves above illustrate the brightness temperature ($T_b$) measured relative to the radio background. The dashed black curve corresponds to a model with standard cosmic expansion/cooling and Pop II star formation (e.g., Mirocha et al. 2017). The red curve is a new model (Mirocha & Furlanetto 2019) consistent with high-z luminosity functions from HST and the recent EDGES results (Bowman et al. 2018a), including added hydrogen cooling motivated by dark matter scattering (e.g., Barkana 2018). Observations at <40 MHz probe the Dark Ages where deviations from the standard cosmological model can be clearly distinguished.

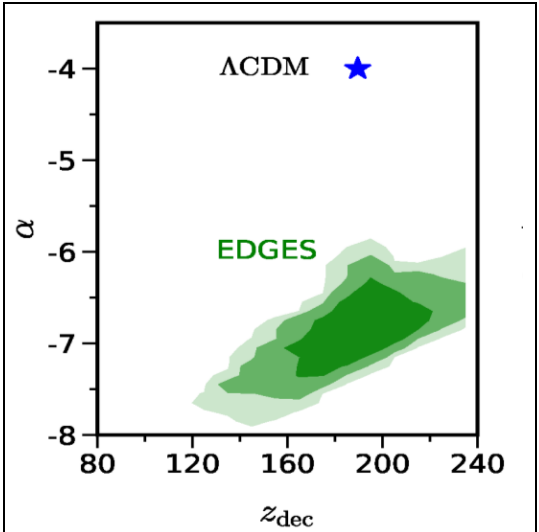
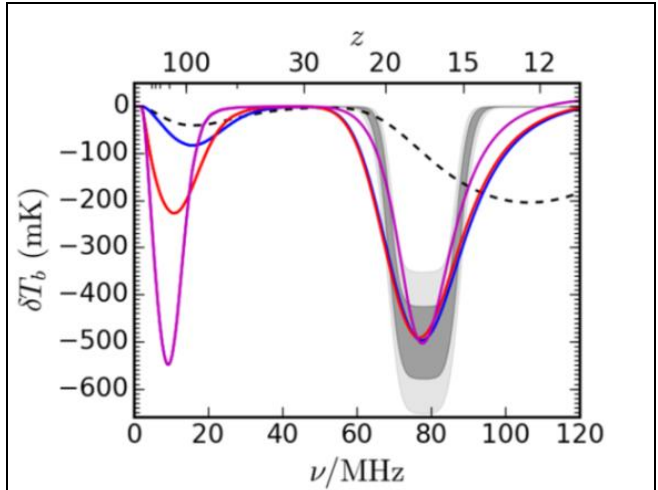

**Figure 3.** Redshifted 21-cm global spectral observations at 15-40 MHz can distinguish at >5σ level between the standard cosmological model whose parameter values are marked by the blue star and models with excess cooling (green contours) here determined by EDGES. These constraints require 15 mK uncertainties across a 15-40 MHz band. The contours represent 1-3σ confidence ranges. The green bands in Figure 6 correspond to the 1σ constraints shown by the dark green contour above.

**Figure 4.** Our parametric models consistent with new EDGES results (grey bands with 1 and 2σ uncertainties) predict a deep absorption trough at <40 MHz produced by additional cooling of baryons, possibly resulting from interaction with dark matter. The black dashed curve (standard cosmology model) is the same as in Figure 2. The blue curve assumes cooling at the adiabatic rate but earlier than the standard model. The red curve has a lower and earlier cooling rate. The magenta curve assumes that the cooling rate does not monotonically decline but rather there is a preferred epoch of excess cooling.

Figure 4 compares the EDGES signal to a model (black dashed curve) bounded by adiabatic expansion and assuming star formation similar to nearby Pop II stars (see e.g., Burns et al. 2017; Mirocha et al. 2017). While the difference in redshift can be explained within the standard model, the trough is about 3 times deeper than expected, suggesting novel physics.

There are three possible explanations for the deep Cosmic Dawn trough in Figures 2 and 4. First, it might be explained by an increase in the radio background ($T_R$ in Eq. 1; e.g., Feng & Holder 2018; Fialkov & Barkana 2019), sourced by synchrotron emission in the first star-forming galaxies or AGNs (e.g., Mirocha & Furlanetto 2019; Ewall-Wice et al. 2018) or dark matter annihilation (Fraser et al. 2018). Second, the trough could also be explained by a change in the cosmological parameters (e.g., $\Omega_{b,0}h^2$ in Eq. 1). A third possibility, which has received much attention, is that the trough could be produced through enhanced cooling of the hydrogen ($T_S$ in Eq. 1) via Rutherford scattering off of dark matter (e.g., Muñoz et al. 2015; Barkana 2018; Fialkov et al. 2018). However, independent constraints suggest that this source of scattering could not compose all of the dark matter in the Universe but rather only a sub-percent fraction (Muñoz & Loeb 2018; Berlin et al. 2018; Kovetz et al. 2018). Identifying excess cooling in the early Universe could thus provide the first evidence that there is more than one kind of dark matter. Indeed, the timing of the signal alone can constrain the properties of any warm component of dark matter (e.g., Safarzadeh et al. 2018; Schneider 2018; Lidz & Hui 2018).

In exploring the implications of the excess baryonic cooling, we follow Mirocha & Furlanetto (2019) and model the cooling rate as

$$\frac{d\log T}{d\log t} = \frac{\alpha}{3} - \frac{(2+\alpha)}{3}\left\{1 + \exp\left[-\left(\frac{z}{z_0}\right)^\beta\right]\right\} \quad \text{Eq. (2)}$$

which can be integrated to obtain the thermal history T(z) of the early Universe. This equation was constructed to reproduce the standard cosmological thermal history when α=-4, and allows a smooth transition between the asymptotic limits of -2/3 when $z \to \infty$ during the epoch of recombination to -4/3 at low-z in the matter-dominated era. The parameter $z_0$ or $z_{dec}$ controls the redshift at which the cooling rate is decoupled halfway between these two limits, β indicates how rapidly the cooling rate declines after Compton scattering becomes inefficient, and α regulates the late-time cooling rate. Fits to the EDGES signal, using the parametric cooling model of Eq. 2, deviate strongly from the ΛCDM prediction at lower frequencies (15-40 MHz), as shown in Figure 3. The blue star shows the position in this plane corresponding to the standard ΛCDM cosmology while the green contours show the expected ranges assuming the true signal matches a cooling model obtained when fitting to the reported EDGES results. It is clear from this figure that a measurement in the 15-40 MHz band will be able to differentiate with high confidence (>5σ) between the presence of added cooling as currently implied by EDGES (see e.g. Barkana 2018; Berlin et al. 2018; Ewall-Wice et al. 2018; Feng & Holder 2018; Fialkov et al. 2018; Fraser et al. 2018; Muñoz & Loeb 2018) and the minimal cosmological cooling.

In Figure 4, we show three example "excess cooling" models that, while consistent with the EDGES Cosmic Dawn trough at 78 MHz, make different predictions for the Dark Ages signal. The dashed black curve again illustrates the depth of the trough predicted using simple adiabatic cooling of the gas plus the effects of Pop II stars as expected before EDGES. The color curves demonstrate the effects of both different cooling rates and time of the cooling, while also adjusting astrophysical parameters in order to preserve the 78 MHz feature seen by EDGES. The 78 MHz Cosmic Dawn trough, while suggestive of exotic physics such as dark matter interactions, is complicated because of the multifaceted astrophysics including star formation, ionization, and black hole X-ray heating that occur during this epoch. The Dark Ages absorption feature described here, reflecting the state of the Universe before the formation of the first stars, has the potential to resolve these ambiguities and cleanly constrain the origin and characteristics of any source of additional cooling. Thus, **the Dark Ages, probed by the redshifted 21-cm signal, is the ideal epoch for a new rigorous test of the standard ΛCDM cosmological model. Divergences from that model would indicate new physics, such as dark matter decay (heating) or baryonic cooling beyond that expected from adiabatic expansion of the Universe.**

### Observations from the Lunar Farside

Observations through the Earth's ionosphere at low radio frequencies, especially below ∼30 MHz, are vitiated by absorption, emission, and refraction at levels well above the expected redshifted HI features (e.g., Datta et al. 2016; Vedantham & Koopmans 2015). Thus, observations above the Earth's ionosphere are necessary in order to avoid corruption of the 21-cm spectrum.

Furthermore, the low radio frequency spectrum is heavily used by powerful civil and military transmitters, with a variety of spectral emissions seen from Earth orbit (see e.g., Burns et al. 2017). To mitigate the effects of this interference, sensitive hydrogen cosmology observations must be conducted from the radio-quiet lunar farside in regions with ∼80 dB suppression of Earth-based RFI (corresponding to ∼mK noise levels). Such observations can be performed from a stable ("frozen") low lunar orbit (50✕125 km; see Plice, Galal & Burns 2017) with data taken when the spacecraft is above the farside or from the farside lunar surface (e.g., Lazio et al. 2011).

### Feasibility and Recommendations

The lunar-based observations described above can be made by a relatively simple instrument with significant flight heritage and the following characteristics: (1) thin, rotating, dual orthogonal dipole antennas which measure linear polarization, (2) low-noise preamplifiers, (3) a dual channel baseband radio spectrometer/polarimeter with 100 MHz of RF bandwidth and onboard cross-spectral capability.

The major challenge in doing hydrogen cosmological observations is the presence of bright galactic and extragalactic foregrounds that are >$10^4$ times that of the 21-cm signal. Our strategy to separate the 21-cm signal takes advantage of three distinct characteristics of the foreground in contrast to the signal. First, as opposed to the broad frequency structure of the 21-cm signal, the frequency spectrum of the foreground is a simple, low-order polynomial (in $\ln \nu - \ln T$ space) produced by synchrotron radiation (e.g., Bernardi et al. 2015; Sathyanarayana Rao et al. 2017) — see Figure 5.

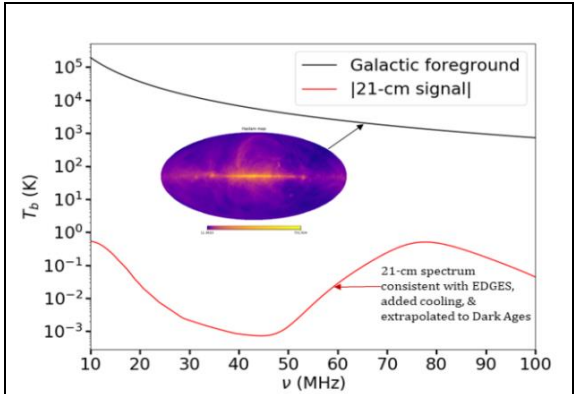

**Figure 5.** The 21-cm spectrum must be measured in the presence of bright spectrally featureless foregrounds. The foreground spectrum (black curve) is shown for a region away from the Galactic center. The red curve illustrates one model of the 21-cm spectrum consistent with EDGES (Figure 4). Note the y-axis is in log($T_b$) which accounts for the difference in shape compared to Figure 4. *Inset:* Mollweide projection of the sky at 408 MHz (Haslam et al. 1982).

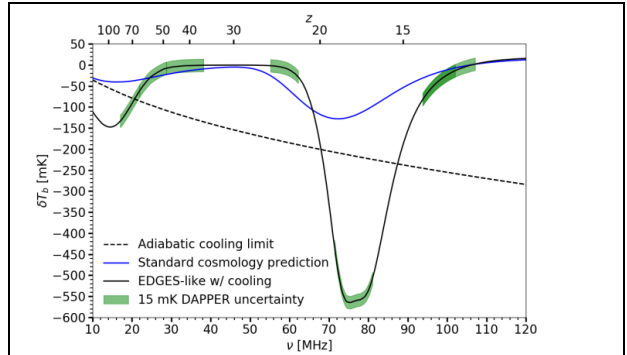

**Figure 6.** The expected performance for a crossed, dual-dipole antenna + spectrometer/polarimeter above or on the lunar farside. Although the primary band is 15-40 MHz, we can use secondary antenna resonances to sparsely sample frequencies up to $\approx$100 MHz. The blue curve is the standard ΛCDM model. The black curve is a model consistent with EDGES including added hydrogen cooling (Figure 4). The dashed curve is the adiabatic gas cooling limit. The green bands show 15 mK uncertainties. Measurements in the 15-40 MHz band will separate the standard cosmology from the EDGES added cooling models at >5$\sigma$.

Second, the Galaxy has significant spatial structure on the sky whereas the cosmological signal is isotropic on scales $\gtrsim$10°. Third, in addition to intrinsic sky polarization, spatial structure in the foreground induces a polarization response in the dipole antenna, whereas the uniform, unpolarized 21-cm signal produces a response only in Stokes I. This, then, allows a clean separation of the primordial signal from the foreground (Burns et al. 2017; Tauscher et al. 2018).

We can take advantage of these differences between the foreground and signal using a new approach to measuring the hydrogen signature. Based upon a single antenna concept, this strategy incorporates two unique features that greatly enhance the probability of a detection. These include (1) dynamic polarimetry, a method for separating the foreground spectrum from that of the HI (Nhan, Bradley & Burns 2017; Nhan et al. 2019), and (2) application of an advanced pattern recognition methodology that characterizes signals and systematics via training sets so that the 21-cm signature can be identified (Tauscher et al. 2018).

We constructed a simulation of a radio frequency spectrometer/polarimeter system that measures the four Stokes parameters to a precision of RMS 15 mK averaged over the band. This simulation makes use of 7-m tip-to-tip thin-wire, rotating dipole antennas that have flown many times on missions such as Wind/WAVES (Bougeret et al. 1995) and the radio frequency FIELDS spectrometer currently in operation on the Parker Solar Probe (Bale et al. 2016; Pulupa et al. 2017). Figure 6 illustrates the expected spectral performance of this instrument. The green uncertainty (1$\sigma$) bands in Figure 6 depict foreground noise corresponding to a total of $\approx$5000 hours of integration over three different antenna deployments, which target different sub-bands. This observational strategy, along with monitoring instrumental systematics and accounting for similarities in the spectral shapes of the foregrounds relative to the 21-cm feature, will allow us to detect deviations at >5$\sigma$ from the standard cosmological model such as those implied by the EDGES results. Thus, we can constrain exotic physics in the Dark Ages for the first time.

The coming decade offers a unique opportunity for RF-quiet measurements from the lunar farside. Rideshare access is expected to be ubiquitous in the short-term, but there is a strong possibility that development of lunar assets will compromise the RF-quiet character in the long-term. We recommend seizing this opportunity to resolve fundamental questions about the validity of the ΛCDM model and the nature of the early Universe by capturing the signature of redshifted 21-cm signal from the Dark Ages.

**References**

Abel, T., Bryan, G. L., Norman, M. L. 2002, "The Formation of the First Star in the Universe," *Science*, 295, 93.

Bale, S. D., Goetz, K., Harvey, P. R., et al. 2016, "The FIELDS Instrument Suite for Solar Probe Plus: Measuring the Coronal Plasma and Magnetic Field, Plasma Waves and Turbulence, and Radio Signatures of Solar Transients," *Space Science Reviews*, 204, 49.

Barkana, R. 2018, "Possible interaction between baryons and dark-matter particles by the first stars," *Nature*, 555, 71.

Behroozi, P. S., Silk, J. 2015, "A Simple Technique for Predicting High-Redshift Galaxy Evolution," *The Astrophysical Journal*, 799, 32.

Berlin, A., Hooper, D., Krnjaic, G., McDermott, S. D. 2018, "Severely Constraining Dark Matter Interpretations of the 21-cm Anomaly," *Physical Review Letters*, 121, 011102.

Bernardi, G., McQuinn, M., Greenhill, L. J. 2015, "Foreground Model and Antenna Calibration Errors in the Measurement of the Sky-averaged λ21 cm Signal at z~ 20," *The Astrophysical Journal*, 799, 90.

Bougeret, J. L., Kaiser, M. L., Kellogg, P. J., et al. 1995, "WAVES: The radio and plasma wave investigation on the wind spacecraft," *Space Science Reviews*, 71, 231.

Bouwens, R. J., Illingworth, G. D., Oesch, P. A., Caruana, J., Holwerda, B., Smit, R., Wilkins, S. 2015, "Reionization After Planck: The Derived Growth of the Cosmic Ionizing Emissivity Now Matches the Growth of the Galaxy UV Luminosity Density," *The Astrophysical Journal*, 811, 140.

Bowman, J. D., et al. 2018a, "An absorption profile centered at 78 MHz in the sky-averaged spectrum," *Nature*, 555, 67.

Bowman, J. D., et al. 2018b, "Reply to Hills et al.," *Nature: Brief Communications Arising*, 564, E35.

Burns, J.O., et al. 2017, "A Space-based Observational Strategy for Characterizing the First Stars and Galaxies using the Redshifted 21-cm Global Spectrum," *The Astrophysical Journal*, 844, 33.

Datta, A., et al. 2016, "The Effects of the Ionosphere on Ground-Based Detection of the Global 21-cm Signal From the Cosmic Dawn and the Dark Ages," *The Astrophysical Journal*, 831, 6.

DeBoer, D. R., et al. 2017, "Hydrogen Epoch of Reionization (HERA)," *Publications of the Astronomical Society of the Pacific*, 129, 045001.

Ewall-Wice, A., Hewitt, J., Mesinger, A., Dillion, J. S., Liu, A., Pober, J. 2016, "Constraining high-redshift X-ray sources with next generation 21-cm power spectrum measurements," *Monthly Notices of the Royal Astronomical Society*, 458, 2710.

Ewall-Wice, A., Chang, T. C., Lazio, J., Doré, O., Seiffert, M., Monsalve, R. A. 2018, "Modeling the Radio Background from the First Black Holes at Cosmic Dawn: Implications for the 21-cm Absorption Amplitude," *The Astrophysical Journal*, 868, 63.

Feng, C., Holder, G. 2018, "Enhanced global signal of neutral hydrogen due to excess radiation at cosmic dawn," *The Astrophysical Journal Letters*, 858, L17.

Fialkov, A., Barkana, R., Cohen, A. 2018, "Constraining Baryon–Dark Matter Scattering with the Cosmic Dawn 21-cm Signal," *Physical Review Letters*, 121, 011101.

Fialkov, A., Barkana, R. 2019, "Signature of Excess Radio Background in the 21-cm Global Signal and Power Spectrum," submitted to the *Monthly Notices of the Royal Astronomical Society*, arXiv:1902.02438.

Fraser, S., Hektor, A., Hütsi, G., Kannike, K., Marzo, C., Marzola, L., Spethmann, C., Racioppi, A., Raidal, M., Vaskonen, V., Veermäe, H. 2018, "The EDGES 21 cm Anomaly and Properties of Dark Matter," *Physics Letters B*, 785, 159.

Furlanetto, S. R., Oh, S. P., Briggs, F. H. 2006, "Cosmology at low frequencies: The 21cm transition and the high-redshift Universe," *Physics Reports*, 433, 181.

Haslam, G., Wielebinski, R., Priester, W. 1982, "Radio Maps of the Sky," *Sky and Telescope*, 63, 230.

Kovetz, E. D., Poulin, V., Glusevic, V., Boddy, K. K., Barkana, R., Kamionkowski, M. 2018, "Tighter Limits on dark matter explanations of the anomalous EDGES 21-cm signal," *Physical Review D*, 98, 3529.

Lazio, T. J. W., MacDowall, R. J., Burns, J. O., et al. 2011, "The Radio Observatory on the Lunar Surface for Solar Studies," *Advances in Space Research*, 48, 1942.

Lidz, A., Hui, L. 2018, "The Implications of a Pre-reionization 21-cm Absorption Signal for Fuzzy Dark Matter," *Physical Review D*, 98, 023011.

Loeb, A., Furlanetto, S. R. 2013, "The First Galaxies in the Universe," ISBN: 9780691144917, Princeton, NJ: Princeton University Press.

McGreer, I. D., Mesinger, A., 'Odorico, V. 2015, "Model-independent evidence in favour of an end to reionization by $z \approx 6$", *Monthly Notices of the Royal Astronomical Society*, 447, 499.

Mesinger, A., McQuinn, M., Spergel, D. N. 2012, "The kinetic Sunyaev-Zel'dovich signal from inhomogeneous reionization: a parameter space study," *Monthly Notices of the Royal Astronomical Society*, 422, 1403.

Mirocha, J. 2014, "Decoding the X-ray Properties of Pre-Reionization Era Sources," *Monthly Notices of the Royal Astronomical Society*, 443, 1211.

Mirocha, J., Furlanetto, S. R., Sun, G. 2017, "The Global 21-cm Signal in the Context of the High-z Galaxy Luminosity Function", *Monthly Notices of the Royal Astronomical Society*, 464, 1365.

Mirocha, J., Furlanetto, S. R. 2019, "What does the first highly-redshifted 21-cm detection tell us about early galaxies?", *Monthly Notices of the Royal Astronomical Society*, 483,1980.

Muñoz, J. B., Kovetz, E. D., Ali-Haimoud, Y. 2015, "Heating of Baryons due to Scattering with Dark Matter During the Dark Ages", *Physical Review Letters*, 92, 083528.

Muñoz, J. B., Loeb, A. 2018, "A small amount of mini-charged dark matter could cool baryons in the early Universe," *Nature*, 557, 684.

Muñoz, J. B., Dvorkin, C., Loeb, A. 2018, "21-cm Fluctuations from Charged Dark Matter", *Physical Review Letters*, 121, 121301.

Nhan, B., Bradley, R., Burns, J. O. 2017, "A Polarimetric Approach for Constraining the Dynamic Foreground Spectrum for Cosmological Global 21 cm Measurements", *The Astrophysical Journal*, 836, 90.

Nhan, B., Bordenave, D., Bradley, R., Burns, J. O., et al. 2019, "A Proof of Concept Experiment to Constrain the Foreground Spectrum for Global 21 cm Cosmology through Projection-Induced Polarimetry", submitted to *The Astrophysical Journal*, arXiv:1811.04917.

Plice, L., Galal, K., Burns, J. O. 2017, "DARE Mission Design: Low RFI Observations from a Low-Altitude Frozen Lunar Orbit," conference proceedings for the 27th AAS/AIAA Space Flight Mechanics Meeting, San Antonio, Texas, arXiv:1702.00286.

Pritchard, J., Loeb, A. 2012, "21-cm cosmology in the 21st Century," *Reports on Progress in Physics*, 086901, 35.

Pulupa, M., et al. 2017, "The Solar Probe Plus Radio Frequency Spectrometer: Measurement requirements, analog design, and digital signal processing," *Journal of Geophysical Resesarch: Space Physics*, 122, 2836.

Safarzadeh, M., Scannapieco, E., Babul, A. 2018, "A Limit on the Warm Dark Matter Particle Mass from the Redshifted 21-cm Absorption Line," *The Astrophysical Journal Letters*, 859, L18.

Sathyanarayana Rao, M., Subrahmanyan, R., Udaya, S. N., Chluba, J. 2017, "Modeling the Radio Foreground for Detection of CMB Spectral Distortions from the Cosmic Dawn and the Epoch of Reionization," *The Astrophysical Journal*, 840, 33.

Schneider, A. 2018, "Constraining Non-Cold Dark Matter Models with the Global 21-cm Signal," *Physical Review D*, 98, 063021.

Shaver, P. A., Windhorst, R. A., Madau, P., de Bruyn, A. G. 1999, "Can the Reionization Epoch be Detected as a Global Signature in the Cosmic Background," *Astronomy and Astrophysics*, 345, 380.

Tauscher, K., Rapetti, D., Burns, J. O., Switzer E. 2018, "Global 21-cm signal extraction from foreground and instrumental effects I: Pattern recognition framework for separation using training sets," *The Astrophysical Journal*, 853, 187.

Vedantham, H. K., Koopmans, L. V. 2015, "Scintillation noise in widefield radio interferometry," *Monthly Notices of the Royal Astronomical Society*, 453, 925.